\documentclass[letterpaper, 10 pt, conference]{ieeeconf}  

\IEEEoverridecommandlockouts                              
\usepackage{float,amsmath,amsfonts,amssymb,amsthm,color,graphicx}
\usepackage{graphics} 
\usepackage{booktabs} 
\usepackage{caption, subcaption}
\usepackage{breqn}
\usepackage{array, mathtools}
\usepackage{pgfplots}
\usepackage{siunitx}
\usepackage{algorithm}
\usepackage[noend]{algpseudocode}
\usepackage{tabulary}
\usepackage{stmaryrd} 
\usepackage{varwidth}

\makeatletter
\newsavebox{\@brx}
\newcommand{\llangle}[1][]{\savebox{\@brx}{\(\m@th{#1\langle}\)}%
  \mathopen{\copy\@brx\kern-0.5\wd\@brx\usebox{\@brx}}}
\newcommand{\rrangle}[1][]{\savebox{\@brx}{\(\m@th{#1\rangle}\)}%
  \mathclose{\copy\@brx\kern-0.5\wd\@brx\usebox{\@brx}}}
\makeatother

\usepackage{tikz}
\usepackage{pgfplots}
\usetikzlibrary{decorations.markings}
\usetikzlibrary{patterns}
\usetikzlibrary{arrows}
\usetikzlibrary{shapes}
\usetikzlibrary{fit}
 \usetikzlibrary{calc}
\usetikzlibrary{decorations.pathreplacing}
\usetikzlibrary{arrows,positioning} 
\usetikzlibrary{pgfplots.groupplots}
\usepackage{soul}
\usepackage{thmtools}

\declaretheorem[style=definition,name=Example,qed=$\blacksquare$]{example}

\declaretheorem[style=plain,name=Proposition]{proposition}
\declaretheorem[style=plain,name=Theorem]{theorem}

\DeclareMathOperator*{\argmin}{arg\, min} 

\newcommand{\R}{\mathbb{R}}

\newcommand{\PP}{\mathcal{P}}
\newcommand{\A}{\mathcal{A}}

\title{\LARGE {\bf Performance Analysis and Non-Quadratic Lyapunov Functions for Linear Time-Varying Systems}}

\author{Matthew Abate, Corbin Klett, Samuel Coogan, and Eric Feron
\thanks{This work was supported by the KAUST baseline budget. 
}
\thanks{M. Abate is with the School of Mechanical Engineering and the School of Electrical and Computer Engineering, Georgia Institute of Technology, Atlanta, 30332, USA: {\tt\small Matt.Abate@GaTech.edu}.}
\thanks{C. Klett is with the School of Aerospace Engineering, Georgia Institute of Technology, Atlanta, 30332, USA: 
{\tt\small Corbin@GaTech.edu}.}
\thanks{S. Coogan is with the School of Electrical and Computer Engineering and the School of Civil and Environmental Engineering, Georgia Institute of Technology, Atlanta, 30332, USA: 
{\tt\small Sam.Coogan@GaTech.edu}.}
\thanks{E. Feron is with the Department of Electrical Engineering, King Abdullah University of Science and Technology, Thuwal, Saudi Arabia: {\tt\small Eric.Feron@Kaust.edu.sa}.}
}

\begin{document}
\def\reals{\mathbb{R}}
\def\Co{{\bf Co}}
\def\diag{{\bf diag}}

\maketitle
\thispagestyle{empty}
\pagestyle{empty}

\begin{abstract}
Performance analysis for linear time-invariant (LTI) systems has been closely tied to quadratic Lyapunov functions ever since it was shown that LTI system stability is equivalent to the existence of such a Lyapunov function. Some metrics for LTI systems, however, have resisted treatment via means of quadratic Lyapunov functions. Among these, point-wise-in-time metrics, such as peak norms, are not captured accurately using these techniques, and this shortcoming has prevented the development of tools to analyze system behavior by means other than \emph{e.g.} time-domain simulations.  This work demonstrates how the more general class of homogeneous polynomial Lyapunov functions can be used to approximate point-wise-in-time behavior for LTI systems with greater accuracy, and we extend this to the case of linear time-varying (LTV) systems as well.  Our findings rely on the recent observation that the search for homogeneous polynomial Lyapunov functions for LTV systems can be recast as a search for quadratic Lyapunov functions for a related hierarchy of time-varying Lyapunov differential equations; thus, performance guarantees for LTV systems are attainable without heavy computation.  Numerous examples are provided to demonstrate the findings of this work.
\end{abstract}

\section{Introduction}
Beginner's courses on linear systems quickly introduce the Lyapunov function as a natural means to express system stability in terms of energy loss. One essential result of Lyapunov states that the stability of a linear time invariant (LTI) system is equivalent to the existence of a quadratic energy function that decays along system trajectories \cite{khalil2002nonlinear}, and since then quadratic stability theory has been greatly extended to develop metrics and indicators of performance such as passivity \cite[Chapter 14]{terrell2009stability}, \cite{boyd1994linear} and robustness \cite{zhou1998essentials}. These metrics generally leverage the ubiquitous presence of the quadratic Lyapunov functions that are naturally embedded in stable LTI systems \cite{hollot1980optimal}. 

Some metrics for LTI systems, however, have resisted treatment via means of quadratic Lyapunov functions. Among these, point-wise-in-time metrics, such as peak norms, are not captured accurately \cite{blanchini1995nonquadratic}, and this shortcoming has prevented the development of tools to analyze system behavior by means other than time-domain simulations. When extending to the case of linear time-varying (LTV) systems, new challenges emerge: for instance, it is known that not all stable LTV systems can be certified via quadratic Lyapunov functions \cite{liberzon2003switching}, and the time-varying nature of these systems reduces the ease of simulation. Further, analytical considerations in simulation are often steered by subjective criteria: for example, the stopping-time of a simulation is, in practice, generally chosen by either analysing the poles of the system or the relative distance to the steady-state output (See, \emph{e.g.} \cite[impulse.m]{MATLAB:2020}).  For these reasons, it is useful to have means other than simulation for extracting time domain properties for LTV systems.

The topic addressed in this paper relies on the recent observation in \cite{abate2019lyapunov} that the search for homogeneous polynomial Lyapunov functions for LTV systems can be recast as the search for quadratic Lyapunov functions for a related hierarchy of Lyapunov differential equations. Indeed, every stable LTV system induces a homogeneous polynomial Lyapunov function \cite{mason2006common, ahmadi2011converse}, and the search for such a Lyapunov function is easily expressed as sum-of-squares and found by solving a convex, semi-definite feasibility program.
Our contribution is to show that the aforementioned hierarchy of LTI systems defines a powerful framework for extracting time-domain properties of LTV systems, and we particularly show how one can compute bounds on the impulse and step response of LTV systems using homogeneous polynomial Lyapunov functions. 

This paper is structured as follows.  We introduce our notation in Section II.   In Section III we recall a procedure for computing norm bounds on the impulse response of LTI systems, and this procedure relies on the use of quadratic Lyapunov functions. In the same section, we introduce a hierarchy of LTI systems that can be used to compute homogeneous polynomial Lyapunov functions for LTI systems.  We show how the aforementioned hierarchy is used to compute bounds on the impulse responses of LTI systems in Section IV and bounds on the step responses of LTI systems in Section V.  Similar bounds are computed for LTV systems in Section VI, and we additionally present a procedure for computing convergence envelopes on the impulse response of LTV systems.  We demonstrate our findings through numerous examples that appear throughout the work and through a case study presented in Section VII.

\section{Notation}
We denote by $S_{++}^n \subset \R^{n\times n}$ the set of symmetric positive definite $n\times n$ matrices. We denote by $I_n$ the $n\times n$ identity matrix, and we denote by $0_n \in \R^n$ the zero vector in $\R^n$.  Given $A \in \R^{n\times m}$ and integer $i \geq 1$, we denote by $\otimes^i A \in \R^{n^i \times m^i}$ the $i^{\rm th}$-Kronecker Power of $A$, as defined recursively by 
\begin{equation}
\begin{split}
    \otimes^1 A &:= A \\
    \otimes^i A &:= A \otimes (\otimes^{i-1} A) \qquad i \geq 2.
\end{split}    
\end{equation}

\section{Preliminaries}
We consider the linear time-invariant system
\begin{equation}\label{LTI_IO}
    \begin{split}
    \dot{x} & = Ax + bu, \\
         y & = c x,
    \end{split}
\end{equation}
with state $x \in \R^n$, control input $u \in \R$ and output $y \in \R$.  We are particularly interested in studying the impulse response of \eqref{LTI_IO}, which is given by 
\begin{equation}\label{impulse}
    h(t) = ce^{At}b.
\end{equation}
Elementary simulations may provide desired information such as a norm-bound on $h(t)$. However, such simulations become cumbersome and inelegant when \emph{e.g.} $A$, $b$, or $c$ are uncertain or time-varying. To address these robustness issues, algebraic approaches to time-domain analyses have been proposed that rely on quadratic Lyapunov functions \cite[Section 6.2]{boyd1994linear}, \cite{abedor1996linear}. 

A quadratic Lyapunov function for \eqref{LTI_IO} is given by $V(x) = x^T P x$ where $P \in S_{++}^{n}$ satisfies 
\begin{equation}\label{lyap_cond}
    A^{T}P + PA \preceq 0.
\end{equation}
Indeed, any quadratic Lyapunov function for \eqref{LTI_IO} implicitly defines an ellipsoidal sublevel set
\begin{equation}\label{ellipse}
{\cal E}_{\alpha} = \left\{x \in \reals^n \mbox{ and } x^TPx\leq \alpha\right\}
\end{equation}
for any positive $\alpha$, and this sublevel set is invariant  in the sense that any trajectory of $\dot{x} = Ax$ that starts within ${\cal E}_\alpha$ stays within ${\cal E}_\alpha$ for all time. Based on this consideration, an upper bound on the impulse response may be obtained from any invariant ellipsoid, as we show in Proposition \ref{prop1}.

\begin{proposition}\label{prop1} \cite{boyd1994linear}
If $P\in S_{++}^n$ satisfies \eqref{lyap_cond}, then $|h(t)| \leq \overline{h}$ for all $t \geq 0$ where 
\begin{equation}\label{corbin}
    \overline{h} = \sqrt{cP^{-1}c^T}\sqrt{b^TPb}.
\end{equation}
\end{proposition}
\begin{proof}
Let $\alpha = b^TPb$.  Then a norm bound on $h(t)$ can be computed by finding the point on the boundary of $\mathcal{E}_{\alpha}$ in the direction $c$; that is $|h(t)| \leq \overline{h}$ for all $t$, where $\overline{h}$ is given by
\begin{equation}\label{maxcx}
    \begin{array}{rcl}
    \overline{h} = & \max\limits_{z\in \R^n} & cz \vspace{.1in}\\
    & \mbox{s.t.} & z^TPz \leq b^TPb\\
    \end{array}
\end{equation}
and this optimization problem is solved by \eqref{corbin}.
\end{proof}

To find the ellipsoid parameter $P$ which minimizes the bound on the impulse response while satisfying the Lyapunov constraint \eqref{lyap_cond}, we formulate the program
\begin{equation}
    \begin{array}{rcl}
    P = & \argmin\limits_{Q\in S_{++}^{n}} & cQ^{-1}c^T \vspace{.1in}\\
    & \mbox{s.t.} & b^TQb \leq 1\\
    &  & A^TQ+PQ \preceq 0
    \end{array}
    \label{convexopt}
\end{equation}
which can be easily computed via convex optimization techniques (See Example \ref{example1})\footnote{In Example \ref{example1}, and those that follow, we compute \eqref{convexopt} using CVX \cite{cvx}, a convex optimization toolbox, made for use with MATLAB. The code that generates the figures from these examples is publicly available through the GaTech FactsLab GitHub: https://github.com/gtfactslab/Abate\_ACC2021}.

\begin{figure}[t]
    \begin{subfigure}{0.48\textwidth}
        \input{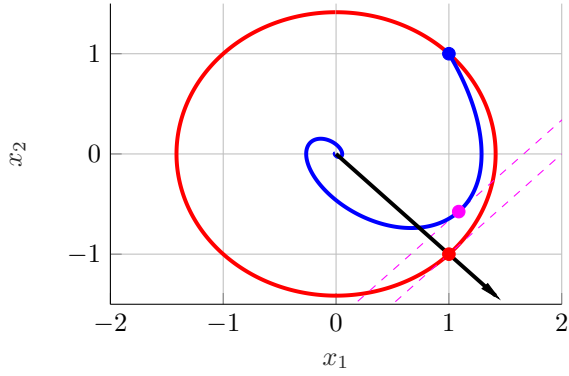}
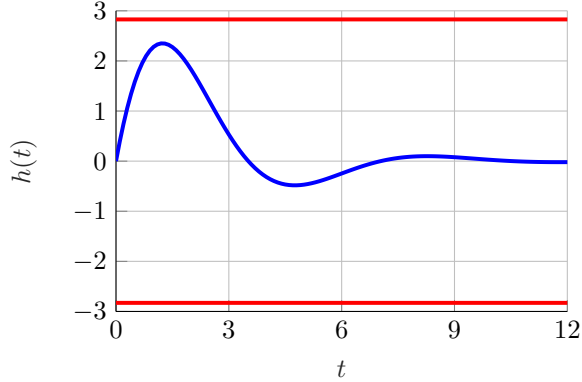
        \caption{Phase portrait of the impulse response for the system \eqref{matt}.
        The impulse response of \eqref{matt} is shown in blue, and $x(0) = b$ is shown as a blue dot. The vector $c^{T}$ is shown with a black arrow, and the state along $x(t)$ which maximizes $h(t)$ is shown as a pink dot.  The invariant ellipse $\mathcal{E}_{1}$ is shown in red and the vector $z$ which solves \eqref{maxcx} is shown as a red dot. The two dashed lines depict a ``gap" of conservatism between the bound produced by $\mathcal{E}_{1}$ and the actual maximum impulse response.\vspace{.5cm}
        }
        \label{fig_impulse}
    \end{subfigure}
    ~
    \begin{subfigure}{0.48\textwidth}
%
%
\begin{tikzpicture}

\begin{axis}[%
width=6cm,
height=4cm,
at={(0cm,0cm)},
scale only axis,
xmin=0,
xmax=12,
xtick={ 0,  3,  6,  9, 12},
xlabel style={font=\color{white!15!black}},
xlabel={$t$},
ymin=-3,
ymax=3,
ytick={-3, -2, -1,  0,  1,  2,  3},
ylabel style={font=\color{white!15!black}},
ylabel={$h(t)$},
axis background/.style={fill=white},
axis x line*=bottom,
axis y line*=left,
xmajorgrids,
ymajorgrids
]
\addplot [color=red, line width=1.5pt, forget plot]
  table[row sep=crcr]{%
0	2.82842712474619\\
12	2.82842712474619\\
};
\addplot [color=red, line width=1.5pt, forget plot]
  table[row sep=crcr]{%
0	-2.82842712474619\\
12	-2.82842712474619\\
};
\addplot [color=blue, line width=1.5pt, forget plot]
  table[row sep=crcr]{%
0	0\\
0.1	0.391554609307157\\
0.2	0.745667180777942\\
0.3	1.06217667702733\\
0.4	1.34129522722644\\
0.5	1.58357398215509\\
0.6	1.78986866863795\\
0.7	1.96130519837358\\
0.8	2.09924565509055\\
0.9	2.20525495246922\\
1	2.28106842368862\\
1.1	2.32856057210405\\
1.2	2.34971518171298\\
1.3	2.34659695597535\\
1.4	2.32132482444249\\
1.5	2.27604702870965\\
1.6	2.21291807260766\\
1.7	2.13407759642711\\
1.8	2.04163121143758\\
1.9	1.93763330911255\\
2	1.82407183936105\\
2.1	1.70285503374378\\
2.2	1.57580003313389\\
2.3	1.44462336457453\\
2.4	1.31093319917093\\
2.5	1.17622331170302\\
2.6	1.04186865321094\\
2.7	0.909122440031801\\
2.8	0.779114656583375\\
2.9	0.652851864521492\\
3	0.531218207656668\\
3.1	0.41497750011017\\
3.2	0.304776284523518\\
3.3	0.201147747608214\\
3.4	0.104516381831395\\
3.5	0.0152032844749008\\
3.6	-0.0665680114239983\\
3.7	-0.140665275744643\\
3.8	-0.207040986439187\\
3.9	-0.265725492893551\\
4	-0.316820023629876\\
4.1	-0.360489618560857\\
4.2	-0.396956059823668\\
4.3	-0.42649086886329\\
4.4	-0.449408430977746\\
4.5	-0.466059302051351\\
4.6	-0.476823745750999\\
4.7	-0.48210554310344\\
4.8	-0.482326110160974\\
4.9	-0.477918953445859\\
5	-0.469324487080744\\
5.1	-0.456985229998805\\
5.2	-0.441341396412334\\
5.3	-0.422826887826635\\
5.4	-0.401865690335935\\
5.5	-0.378868676743852\\
5.6	-0.354230809222164\\
5.7	-0.328328734763254\\
5.8	-0.301518762594979\\
5.9	-0.274135210009182\\
6	-0.246489100700928\\
6.1	-0.218867197715953\\
6.2	-0.191531351447402\\
6.3	-0.164718141796005\\
6.4	-0.138638792594609\\
6.5	-0.113479335681311\\
6.6	-0.0894010015664634\\
6.7	-0.066540813457975\\
6.8	-0.0450123614660457\\
6.9	-0.024906734081799\\
7	-0.00629358449293105\\
7.1	0.0107776900577768\\
7.2	0.0262766768696038\\
7.3	0.0401906044073895\\
7.4	0.0525229459823568\\
7.5	0.063291988389129\\
7.6	0.0725293821921338\\
7.7	0.0802786890931623\\
7.8	0.0865939405115488\\
7.9	0.0915382201850301\\
8	0.0951822822674235\\
8.1	0.0976032150723681\\
8.2	0.098883159302662\\
8.3	0.0991080883233156\\
8.4	0.0983666567930603\\
8.5	0.0967491227723871\\
8.6	0.0943463472835865\\
8.7	0.0912488742160255\\
8.8	0.087546092453169\\
8.9	0.083325481150689\\
9	0.0786719382204908\\
9.1	0.0736671912757183\\
9.2	0.0683892895679761\\
9.3	0.0629121748005268\\
9.4	0.0573053281296806\\
9.5	0.0516334901699355\\
9.6	0.0459564503949411\\
9.7	0.0403289019737886\\
9.8	0.0348003577977642\\
9.9	0.0294151232333575\\
10	0.0242123209795127\\
10.1	0.0192259633070601\\
10.2	0.0144850669119293\\
10.3	0.0100138056169828\\
10.4	0.00583169620577683\\
10.5	0.00195381276094076\\
10.6	-0.00160897499423696\\
10.7	-0.00484974369435523\\
10.8	-0.00776524282794553\\
10.9	-0.0103556096786907\\
11	-0.0126240764301346\\
11.1	-0.0145766729070422\\
11.2	-0.016221928169337\\
11.3	-0.0175705739087123\\
11.4	-0.0186352523268786\\
11.5	-0.0194302309010092\\
11.6	-0.0199711261690959\\
11.7	-0.0202746383981765\\
11.8	-0.0203582987340673\\
11.9	-0.0202402301743748\\
12	-0.0199389234589947\\
};
\end{axis}
\end{tikzpicture}%
        \caption{Impulse response for the system \eqref{matt}, plotted in the time domain.  The impulse response of \eqref{matt} is shown in blue, and the magnitude bound $\overline{h} = 2\sqrt{2}$ is shown in red.}
		\label{fig_impulse2}
    \end{subfigure}
    \caption{  
    Example 1. Figures \ref{fig_impulse} and \ref{fig_impulse2} plot the impulse response of the system \eqref{matt} in the phase plane and time domain, respectively.
    }
    \label{fig1}
\end{figure}

\begin{example}\label{example1}
Consider the system
\begin{equation}\label{matt}
    \dot{x} = 
    \begin{bmatrix}
    0 & 1 \\ -1 & -0.9
    \end{bmatrix}x + 
    \begin{bmatrix}
    1 \\ 1
    \end{bmatrix}
    u,
\end{equation}
\begin{equation*}
    y = \begin{bmatrix}
    \sqrt{2} & -\sqrt{2}
    \end{bmatrix} x.
\end{equation*}
Solving \eqref{convexopt}, we have that $P = (0.5)I_2$ is the ellipsoidal parameter that minimizes impulse response bound given in \eqref{corbin}.  Then, from \eqref{corbin} we find $|h(t)|\:\, \leq \overline{h} = 2\sqrt{2}$  for all $t$. Figures \ref{fig_impulse} and \ref{fig_impulse2} plot the impulse response of \eqref{matt} in the phase plane and time domain, respectively.
\end{example}

The maximum impulse response of a passive system is known to be equal to $\overline{h}$ from \eqref{corbin} \cite{boyd1994linear}, however, the norm bound \eqref{corbin} generally suffers from conservatism (See Example \ref{example2}).

\begin{example}\label{example2}
Consider the system \eqref{LTI_IO} with stiff dynamics 
\begin{equation*}
\begin{split}
    (A)_{i,\, j} &= 
\begin{cases}
-(M)^{i-1} & \text{if $i = j$} \\
0 & \text{otherwise,}
\end{cases}\\
(b)_i &= 1 \\
(c)_i &= 
\begin{cases}
1 & \text{if $i = 1$} \\
(-1)^{i+1}2 & \text{otherwise.}
\end{cases}
\end{split}
\end{equation*}
for some $M \geq 0$. This system is stable, and $|h(t)|\: \leq 1$ for all $t \geq 0$. However, it was shown in \cite{feronThesis} that the gap between the actual maximum impulse response and the upper bound obtained by solving \eqref{convexopt} grows to $2n-1$ when $M$ tends toward infinity. 
\end{example}

In this work, we address the aforementioned conservatism, and we present a similar technique for generating a norm bound on $h(t)$ that relies on nonquadratic Lyapunov functions for the system $\dot{x}= Ax$.  To that end, we first define the following infinite hierarchy of LTI systems:
\begin{equation}\label{hierarchy}
\begin{split}
H_1 : & \begin{cases}
 \dot{\xi}_1 = \A_1\xi_1 \\
 \A_1=A
\end{cases} \\
H_i : & \begin{cases}
 \dot{\xi}_i = \A_i\xi_i \\
 \A_i = I_n \otimes \A_{i-1} + A\otimes I_{n^{i-1}},\; i\geq 2,
\end{cases} 
\end{split}
\end{equation}
where $A\in \R^{n\times n}$ and the state of the system $H_i$ is given by $\xi_i \in \R^{n^i}$.  This hierarchy is best understood by looking at $H_2$, which is the vectorized version of the Lyapunov differential equation $\dot{X} = AX+XA^T$ with $X\in \R^{n\times n}$.  Moreover, if $x(t)$ is a solution to $\dot{x} = A x$ then $\xi_i(t) = (\otimes^i x(t))$ is a solution to $H_i$. This hierarchy is closely tied to the Liouville equations used to obtain the infinite-dimensional linear differential equations which drive the evolution of probability density functions, in a way similar to the Chapman-Kolmogorov equations \cite[Chapter 16]{lieberman2005introduction}. See also \cite{CorbinKlett} for the discrete-time parallel to \eqref{hierarchy}.

An essential observation made in \cite{abate2019lyapunov} is that a quadratic Lyapunov function for the $i^{\text{th}}$ level system $H_i$ identifies a homogeneous polynomial Lyapunov function for $\dot{x} = Ax$; that is, if $\PP_i \in S_{++}^{n^i}$ satisfies
\begin{equation}\label{metalyap_cond}
    \A_i^T\PP_i +\PP_i\A_i \preceq 0
\end{equation}
for $\A_i$ as defined in \eqref{hierarchy}, then a polynomial Lyapunov function for the system $\dot{x} = Ax$ is given by \begin{equation}
    V(x) = (\otimes^i x)^{T}\PP_i (\otimes^i x)
\end{equation}
and this Lyapunov function is homogeneous in the entries of $x$ and is of order $2i$.

\section{Impulse Response Analysis via Homogeneous Polynomial Lyapunov Functions}
Our main result is to show that the impulse response bound on $h(t)$, which is provided in Proposition \ref{prop1}, can be considerably improved when higher-order polynomial Lyapunov functions are considered. In particular, we study the guarantees attainable when considering the homogeneous polynomial Lyapunov functions that naturally arise from the hierarchy of stable LTI systems (\ref{hierarchy}), and we show in the following theorem how these Lyapunov functions are used to bound the impulse response $h(t)$.

For integer $i \geq 1$, define $\mathbf{b}_i \in \R^{n^i}$ and $\mathbf{c}_i^{T} \in \R^{n^i}$ by
\begin{equation*}
        \mathbf{b}_i = \otimes^{i} b, \qquad 
        \mathbf{c}_i = \otimes^{i} c.
\end{equation*}

\begin{theorem}\label{thrm1}
If $\PP_i \in S_{++}^{n^i}$ satisfies \eqref{metalyap_cond} at the $i^{\text th}$ level, then $|h(t)|\; \leq \overline{h}$ for all $t$ where
\begin{equation}\label{newbound}
    \overline{h} = (\mathbf{c}_i\PP^{-1}_{i} \mathbf{c}_i^T)^{1/(2i)}
    (\mathbf{b}_i^T\PP_{i} \mathbf{b}_i)^{1/(2i)}.
\end{equation}
\end{theorem}
\begin{proof}
For any integer $i\geq 1$, construct the system 
\begin{equation}\label{meta2}
    \begin{split}
        \dot{\xi} &= \A_i\xi + \mathbf{b}_i u \\
        y &= \mathbf{c}_i \xi
    \end{split}
\end{equation}
with $\xi \in \R^{n^i}$, $u \in \R$, and impulse response $\mathbf{h}(t)$. Assuming $\PP_i \in S_{++}^{n^i}$ satisfies \eqref{metalyap_cond} at the $i^{\text th}$ level, we have that $|\mathbf{h}(t)|\,\leq (\mathbf{c}_i\PP^{-1}_i\mathbf{c}_i^T)^{1/2}(\mathbf{b}_i^T\PP_i\mathbf{b}_i)^{1/2}$.  Moreover, from the construction \eqref{meta2}, we have that $|h(t)| \leq |\mathbf{h}(t)|^{1/i} = \overline{h}$.  This competes the proof.
\end{proof}

As in \eqref{convexopt}, we next formulate a convex program  to search for the parameter $\PP_i$ which provides the tightest upper bound on $h(t)$ attainable using Theorem \ref{thrm1}:
\begin{equation}
    \begin{array}{rcl}
    \PP_i = & \argmin\limits_{Q\in S_{++}^{n^i}} & {\mathbf{c}_i}Q^{-1}{\mathbf{c}_i}^T \vspace{.1in}\\
    & \mbox{s.t.} & {\mathbf{b}_i}^T Q{\mathbf{b}_i} \leq 1\\
    &  & \A_i^T Q + Q\A_i \preceq 0.
    \end{array}
    \label{LMI}
\end{equation}
We demonstrate the application of Theorem \ref{thrm1} in Example \ref{example3}.

\begin{figure}[t]
    \begin{subfigure}{0.48\textwidth}
        \input{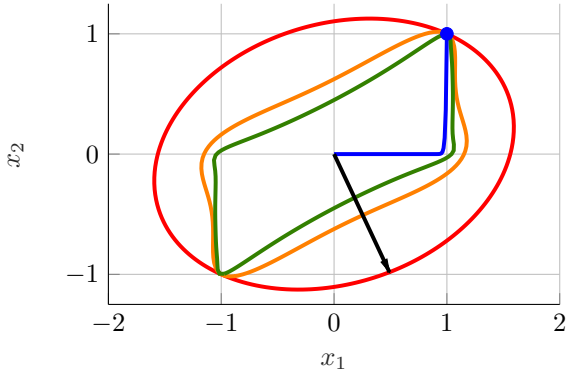}
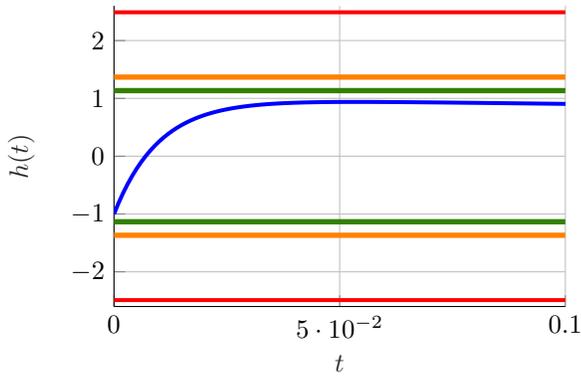
        \caption{
        Phase portrait of the impulse response for the stiff system from Example \ref{example2} where $n = 2$ and $M = 100$. The impulse response is shown in blue, and $x(0) = b$ is shown as a blue dot. A vector in the direction of $c^{T}$ is shown with a black arrow.  The invariant sublevel sets of the $2^{\rm nd},\, 4^{\rm th}$ and $10^{\rm th}$ order homogeneous polynomial Lyapunov functions that are derived in this study are shown in red, orange and green, respectively.\vspace{.5cm}
        }
        \label{impulse3}
    \end{subfigure}
    ~
    \begin{subfigure}{0.48\textwidth}
%
%
\definecolor{mycolor1}{rgb}{0.20000,0.50000,0.00000}%
\begin{tikzpicture}

\begin{axis}[%
width=6cm,
height=4cm,
at={(0cm,0cm)},
scale only axis,
xmin=0,
xmax=0.1,
xtick={   0, 0.05,  0.1},
xlabel style={font=\color{white!15!black}},
xlabel={$t$},
ymin=-2.6,
ymax=2.6,
ytick={-2, -1,  0,  1,  2},
ylabel style={font=\color{white!15!black}},
ylabel={$h(t)$},
axis background/.style={fill=white},
axis x line*=bottom,
axis y line*=left,
xmajorgrids,
ymajorgrids
]
\addplot [color=blue, line width=1.5pt, forget plot]
  table[row sep=crcr]{%
0	-1\\
0.001	-0.810674336238544\\
0.002	-0.639459507488631\\
0.003	-0.484631945860063\\
0.004	-0.344632102727287\\
0.005	-0.218048840232585\\
0.006	-0.103605308134118\\
0.007	-0.000146164649583835\\
0.008	0.0933739866026175\\
0.009	0.177901059291686\\
0.01	0.254290951406283\\
0.011	0.32331811137921\\
0.012	0.385683289037526\\
0.013	0.442020548952262\\
0.014	0.492903616379649\\
0.015	0.538851619306203\\
0.016	0.580334284065974\\
0.017	0.61777663652944\\
0.018	0.651563255915128\\
0.019	0.682042123797536\\
0.02	0.70952810683353\\
0.021	0.734306108063496\\
0.022	0.756633918326542\\
0.023	0.77674479632767\\
0.024	0.794849803179084\\
0.025	0.811139914780535\\
0.026	0.825787933180082\\
0.027	0.838950216044837\\
0.028	0.850768241550811\\
0.029	0.86137002435379\\
0.03	0.87087139681278\\
0.031	0.87937716828891\\
0.032	0.886982174122465\\
0.033	0.893772224786552\\
0.034	0.899824964716854\\
0.035	0.90521064941293\\
0.036	0.909992848588538\\
0.037	0.914229082408375\\
0.038	0.917971397178868\\
0.039	0.921266886254758\\
0.04	0.924158161374855\\
0.041	0.926683779144276\\
0.042	0.928878626931529\\
0.043	0.930774272042629\\
0.044	0.93239927766691\\
0.045	0.933779488756615\\
0.046	0.934938290701448\\
0.047	0.935896843386979\\
0.048	0.936674292979465\\
0.049	0.937287963556656\\
0.05	0.937753530502543\\
0.051	0.938085177401396\\
0.052	0.938295738001368\\
0.053	0.938396824668478\\
0.054	0.938398944616573\\
0.055	0.938311605076556\\
0.056	0.93814340845743\\
0.057	0.937902138451581\\
0.058	0.937594837946047\\
0.059	0.937227879519563\\
0.06	0.936807029230916\\
0.061	0.936337504337038\\
0.062	0.935824025518497\\
0.063	0.935270864135075\\
0.064	0.934681884984382\\
0.065	0.934060584991448\\
0.066	0.933410128216523\\
0.067	0.93273337753143\\
0.068	0.932032923281524\\
0.069	0.931311109220105\\
0.07	0.930570055974839\\
0.071	0.92981168228106\\
0.072	0.929037724194452\\
0.073	0.928249752475405\\
0.074	0.927449188319028\\
0.075	0.926637317588257\\
0.076	0.925815303692501\\
0.077	0.924984199240722\\
0.078	0.92414495658558\\
0.079	0.923298437364169\\
0.08	0.922445421130831\\
0.081	0.92158661316845\\
0.082	0.920722651556404\\
0.083	0.919854113565909\\
0.084	0.918981521446767\\
0.085	0.918105347663436\\
0.086	0.917226019632816\\
0.087	0.916343924011177\\
0.088	0.915459410573135\\
0.089	0.914572795721488\\
0.09	0.913684365663055\\
0.091	0.912794379282286\\
0.092	0.911903070741416\\
0.093	0.911010651833184\\
0.094	0.910117314109654\\
0.095	0.909223230808456\\
0.096	0.908328558595724\\
0.097	0.907433439143166\\
0.098	0.906538000555057\\
0.099	0.905642358659436\\
0.1	0.904746618176435\\
0.101	0.903850873775414\\
0.102	0.902955211031509\\
0.103	0.902059707291145\\
0.104	0.901164432455188\\
0.105	0.900269449687566\\
0.106	0.899374816056461\\
0.107	0.898480583114479\\
0.108	0.897586797423613\\
0.109	0.896693501030244\\
};
\addplot [color=red, line width=1.5pt, forget plot]
  table[row sep=crcr]{%
0	2.48920664489519\\
0.1	2.48920664489519\\
};
\addplot [color=orange, line width=2.0pt, forget plot]
  table[row sep=crcr]{%
0	1.36795669399836\\
0.1	1.36795669399836\\
};
\addplot [color=mycolor1, line width=2.0pt, forget plot]
  table[row sep=crcr]{%
0	1.13395345924129\\
0.1	1.13395345924129\\
};
\addplot [color=red, line width=1.5pt, forget plot]
  table[row sep=crcr]{%
0	-2.48920664489519\\
0.1	-2.48920664489519\\
};
\addplot [color=orange, line width=2.0pt, forget plot]
  table[row sep=crcr]{%
0	-1.36795669399836\\
0.1	-1.36795669399836\\
};
\addplot [color=mycolor1, line width=2.0pt, forget plot]
  table[row sep=crcr]{%
0	-1.13395345924129\\
0.1	-1.13395345924129\\
};
\end{axis}
\end{tikzpicture}%
        \caption{
        Impulse response for the system \eqref{matt}, plotted in the time domain.  The impulse response of \eqref{matt} is shown in blue, and the magnitude bounds derived using $\PP_i$ for $i = 1,\, 2,\, 5$ are shown in red, orange and green, respectively.  As $t$ goes to infinity, $h(t)$ decays to $0$.}
		\label{impulse4}
    \end{subfigure}
    \caption{  
    Example 3. Figures \ref{impulse3} and \ref{impulse4} plot the impulse response of the stiff system from Example \ref{example2}, where $n = 2$ and $M = 100$.
    }
    \label{fig2}
\end{figure}

\begin{example}\label{example3}
We consider the stiff system, previously presented in Example \ref{example2}, where we take $n = 2$ and $M = 100$.  The optimization problem \eqref{LMI} is solved for $i = 1,\, 2,\, 5,$ and the resulting quadratic Lyapunov parameters $\PP_i$ are used to generate bounds on the impulse response using \eqref{newbound}. In Figure \ref{fig2}, we show the impulse response of the stiff system and the bounds derived using Theorem \ref{thrm1}.  Note that as the degree of the Lyapunov functions grows, the sublevel sets of the resulting Lyapunov function shrink and approximate the relevant parts of the impulse response trajectory in the state-space with greater accuracy.
\end{example}

As illustrated in Example \ref{example3}, the accuracy of the bound \eqref{newbound} will generally increase as $i$ increases.  This is due to the fact that the homogeneous polynomial Lyapunov functions generalize quadratic Lyapunov functions \cite{abate2019lyapunov}.

Theorem \ref{thrm1} can also be used to reduce the simulation complexity for systems of the form \eqref{LTI_IO}.  While it is natural to analyse such systems though simulation, it can be ambiguous as when a simulation should be stopped.  In the following example, we demonstrate one potential solution to this problem, whereby a system is simulated over a given amount of time and then the remaining simulation output is bounded using \eqref{newbound}.

\begin{figure}
	\input{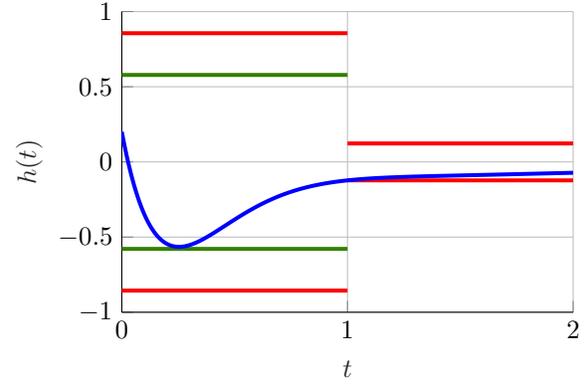}\\
	\caption{Example \ref{example4}. The impulse response $h(t)$ is shown in blue, and the magnitude bounds derived using $\PP_1$ and $\PP_4$ are shown in red and green, respectively. At time $t = 1$, a bound on the tail of $h(t)$ is computed via \eqref{newbound} with $i = 1$, and this bound is shown in red.}
	\label{fig3}
\end{figure}

\begin{example}\label{example4}
Consider the system \eqref{LTI_IO} with $x\in \R^3$ and
\begin{equation*}
    A = \begin{bmatrix}
    0.3 & 0.5 & 10 \\
    -1 & -1.7 & 1 \\
    -2 & -2 & -7.7
    \end{bmatrix}, \hspace{.1in}
    b = \begin{bmatrix}
    0.2 \\ 1 \\ 1
    \end{bmatrix}, \hspace{.1in}
    c^T = \begin{bmatrix}
    1 \\ - 2 \\ 2
    \end{bmatrix}.
\end{equation*}
The optimization problem \eqref{LMI} is solved for $i = 1,\, 4,$ and the resulting Lyapunov parameters $\PP_i$ are used to generate bounds on the impulse response using \eqref{newbound}. At time $t = 1$, a new bound is computed via \eqref{newbound} where $b$ is now taken to be the simulated state $x(1)$; this creates a norm bound on the tail of the impulse response, as shown in Figure \ref{fig3}.
\end{example}

\section{Step Response Analysis via Homogeneous Polynomial Lyapunov Functions}
We next turn our discussion to the step response of \eqref{LTI_IO}, which is the output $y(t)$ when $x(0) = 0_n$ and $u(t) = 1$ for all $t \geq 0$.  Equivalently, the step response of \eqref{LTI_IO} is given by $s(t)$ where 
\begin{equation}\label{step}
\begin{split}
    \dot{x} &= Ax + b \\
    s &= cx
\end{split}
\end{equation}
and $x(0) = 0_n,$ and a closed form representation of the step response is given by
\begin{equation}
    s(t) = c A^{-1}(e^{At} - I_n)b.
\end{equation}

As we show next, a norm bound on $s(t)$ can be derived using Lyapunov functions in manner similar to that presented previously. For integer $i \geq 1$, define $\mathbf{A}_i \in \R^{n^i \times n^i}$ by
\begin{equation*}
        \mathbf{A}_i = \otimes^{i} (A^{-1}).
\end{equation*}

\begin{theorem}\label{thrm2}
If $\PP_i \in S_{++}^{n^i}$ satisfies \eqref{metalyap_cond} at the $i^{\text th}$ level, then $|s(t) + cA^{-1}b|\; \leq \overline{s}$ for all $t$ where
\begin{equation}\label{step_bound}
    \overline{s} = (\mathbf{c}_i\PP^{-1}_i\mathbf{c}_i^T)^{1/(2i)}
    (\mathbf{b}_i^T\mathbf{A}_i^{T}\PP_i\mathbf{A}_i\mathbf{b}_i)^{1/(2i)}.
\end{equation}
\end{theorem}
\begin{proof}
The system \eqref{step} has a stable equilibrium $x_{\rm eq} = -A^{-1}b$. Taking the transformation $\widetilde{x}(t) = x(t) - x_{\rm eq}$, we find that the step response of \eqref{LTI_IO} is equal to the impulse response of the system
\begin{equation}
\begin{split}
    \dot{\widetilde{x}} &= A \widetilde{x} + A^{-1}bu \\
    y &= c\widetilde{x} - cA^{-1}b.
\end{split}
\end{equation}
Thus, the bound \eqref{step_bound} is derived using Theorem \ref{thrm1}.
\end{proof}

Using similar reasoning to that of \eqref{convexopt}, we find that the Lyapunov parameter $\PP_i$ that provides the tightest upper bound on $|s(t) + cA^{-1}b|$ attainable using Theorem \ref{thrm2} is given by
\begin{equation}\label{opt2}
    \begin{array}{rcl}
    \PP = & \argmin\limits_{Q\in S_{++}^{n^i}} & {\mathbf{c}_i}Q^{-1}{\mathbf{c}_i}^T \vspace{.1in}\\
    & \mbox{s.t.} & \mathbf{b}_i^T\mathbf{A}_i^T Q{\mathbf{A}_i\mathbf{b}_i} \leq 1\\
    &  & \A_i^T Q + Q\A_i \preceq 0.
    \end{array}
\end{equation}
We demonstrate the application of Theorem \ref{thrm2} in Example \ref{example5}. 

\begin{figure}
    \input{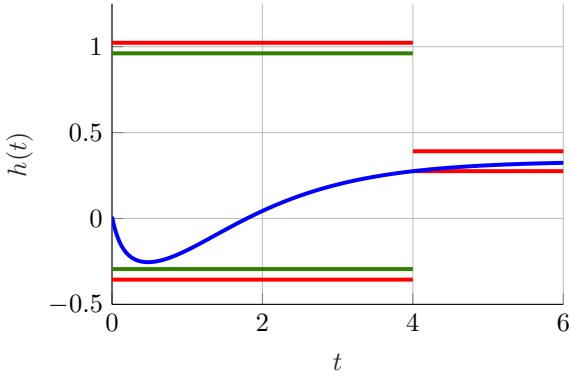}\\
	\caption{Example \ref{example5}. The step response $s(t)$ is shown in blue, and the magnitude bounds derived using the $\PP_i$ for $i = 1,\, 3$ are shown in red and green, respectively. At time $t = 1$, a new bound is computed via \eqref{step_bound} using $i = 1$ and this bound is shown in red.}
    \label{fig4}
\end{figure}

\begin{example}\label{example5}
Consider the system \eqref{LTI_IO} with $x\in \R^3$ and
\begin{equation*}
    A = \begin{bmatrix}
    -1 & 0 & 2 \\
     0 & -10 & 1 \\
     0 & -2 & - 1
    \end{bmatrix}, \hspace{.1in}
    b = \begin{bmatrix}
    -2 \\ 1 \\ 1
    \end{bmatrix}, \hspace{.1in}
    c^T = \begin{bmatrix}
    1 \\ - 2 \\ 2
    \end{bmatrix}.
\end{equation*}
The optimization problem \eqref{opt2} is solved for $i = 1,\, 3,$ and the resulting Lyapunov parameters $\PP_i$ are used to generate bounds on the step response using \eqref{step_bound}. At time $t = 4$, a new bound on the step response is computed via \eqref{step_bound}, and this creates a norm bound on the tail of $s(t)$, as shown in Figure \ref{fig4}.
\end{example}

\section{Impulse Response Analysis for Linear Time-Varying Systems}
The foregoing ideas can be used over a range of possible applications going beyond the analysis of a single LTI system. Lyapunov functions have long been known to be useful for robustness analyses, and we explore these applications in this section.

\subsection{Bounds on Impulse Response for Uncertain and Nonlinear Systems}
The foregoing bounds on impulse response can be readily extended to the case of linear time-varying input-output systems. Specifically, we consider
\begin{equation}
\begin{split}\label{LDI}
\dot{x} & = A(t)x + bu, \\
y &= cx,
\end{split}
\end{equation}
where for given $A,\, \Delta \in \R^{n\times n}$, we have that 
\begin{equation}\label{acond}
    A(t) \in \Big{\{} A + \lambda \Delta \:\vert\: \lambda \in [-1, 1]\,\Big{\}}
\end{equation}
for all $t$.  In this case, the impulse response $h(t)$ is described parametrically by a solution $\varphi(t)$ to
\begin{equation}\label{solution}
\begin{split}
    \dot{\varphi}(t) & = A(t)\varphi(t) \\
    h(t) &= c\varphi(t) \\
    \varphi(0) &= b.
\end{split}
\end{equation}

Theorem \ref{thrm3} shows how a norm bound on $h(t)$ for \eqref{LDI} can be  similarly computed by considering the hierarchy \eqref{hierarchy}.

\begin{theorem}\label{thrm3}
If $\PP_i \in S_{++}^{n^i}$ satisfies \eqref{metalyap_cond} for $A + \Delta$ and $A - \Delta$ at the $i^{\text th}$ level, then $|h(t)|\; \leq \overline{h}$ for all $t$ where $\overline{h}$ is given by \eqref{newbound}.
Moreover, the parameter $\PP_i$ which minimizes $\overline{h}$ can be computed with \eqref{convexopt}, where $\PP_i$ is understood to satisfy \eqref{metalyap_cond} for both $A + \Delta$ and $A - \Delta$ at the $i^{\rm th}$ level.
\end{theorem}
\begin{proof}
Assume there exists a $\PP_i \in S_{++}^{n^i}$ that satisfies \eqref{metalyap_cond} for both $A + \Delta$ and $A - \Delta$ at the $i^{\text th}$ level.  Then, the system $\dot{x} = A(t)x$ is stable with a homogeneous polynomial Lyapunov function $V(x) = (\otimes^i x)^T \PP_i (\otimes^i x)$ \cite{abate2019lyapunov}.  Therefore $|h(t)|\; \leq \overline{h}$, where $h(t)$ is the impulse response of \eqref{LDI} and $\overline{h}$ the a point on the level set $\{x\in \R^n \,\vert\, V(x) = \mathbf{b}_i^T \PP_i \mathbf{b}_i\}$ in the direction $c^T$.  It follows from the reasoning presented in the proof of Theorem \ref{thrm1} that $\overline{h}$ is given by \eqref{newbound}. This completes the proof.
\end{proof}

The stability guarantees in Theorem \ref{thrm3} are in terms of a global norm bound on $h(t)$.  We next generalise Theorem \ref{thrm3} to provide a time-dependent bound on $h(t)$ which is exponentially growing/decaying in $t$ (See Theorem \ref{thrm4}).

\begin{theorem}\label{thrm4}
For $\alpha \in \R$, if $\PP_i \in S_{++}^{n^i}$ satisfies \eqref{metalyap_cond} for $A + \Delta + \alpha I_n$ and $A - \Delta + \alpha I_n$ at the $i^{\text th}$ level, then $|h(t)|\; \leq e^{-\alpha t}\overline{h}$ for all $t$ where $\overline{h}$ is given by \eqref{newbound}.
\end{theorem}
\begin{proof}
Choose $\alpha \in \R$, and assume there exists a $\PP_i \in S_{++}^{n^i}$ that satisfies \eqref{metalyap_cond} for both $A + \Delta + \alpha I_n$ and $A - \Delta + \alpha I_n$ at the $i^{\text th}$ level.  Then, $V(x) = (\otimes^i x)^T \PP_i (\otimes^i x)$ is a homogeneous polynomial Lyapunov function for the system
\begin{equation}
    \dot{x} = A_{\alpha}(t) x.
\end{equation}
where $A_{\alpha}(t)\in \R^{n\times n}$ evolves according to
\begin{equation}
    A_{\alpha}(t) \in \Big{\{} A + \alpha I_n + \lambda \Delta \, \Big{\vert} \, \lambda \in [-1,\, 1] \Big{\}}.
\end{equation}
Thus, applying the results of Theorem \ref{thrm3}, we have that the impulse response of 
\begin{equation}\label{alpha_sys}
    \begin{split}
        \dot{x} &= (A(t) + \alpha I_n)x + bu\\
        y &= c x
    \end{split}
\end{equation}
is bounded by $\overline{h}$ from \eqref{newbound}. 

Fix an $A(t)$ satisfying \eqref{acond}, and denote by $\varphi(t) \in \R^n$, $h(t) \in \R$ the solution to \eqref{solution}.  
Then $\varphi_{\alpha}(t) := e^{\alpha t} \varphi(t)$ is the solution to 
\begin{equation}
    \begin{split}
        \dot{\varphi}_{\alpha}(t) &= (A(t) + \alpha I_n)\varphi_{\alpha}(t) \\
        \varphi_{\alpha}(0) &= b. 
    \end{split}
\end{equation}
Therefore $|h(t)|\; \leq e^{-\alpha t}\overline{h}$ for all $t$.  This completes the proof.
\end{proof}

We demonstrate application of Theorems \ref{thrm3} and \ref{thrm4} in Example \ref{example6}.

\begin{figure}
    \input{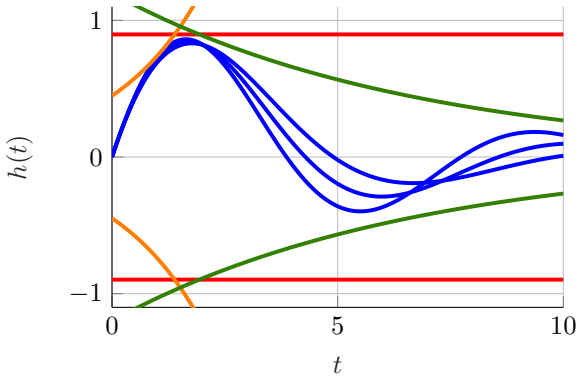}\\
	\caption{Example \ref{example6}. Three system simulations are conducted, and the impulse responses of each is plotted in blue. A global norm bound on $h(t)$ is computed using Theorem \ref{thrm3} for $i=6$, and this bound is shown in red. Two exponential bounds on $h(t)$ are constructed from Theorem \ref{thrm4} with $\alpha = -0.5,\, 0.15$ and $i = 6$, and the bounds are shown in orange and green, respectively.}
    \label{fig5}
\end{figure}

\begin{example}\label{example6}
Consider the uncertain system \eqref{LDI} with $x\in \R^2$, and 
\begin{equation}
\begin{array}{ll}
\begin{split}
A &= \begin{bmatrix} 0 &   1\\ -0.6 &   -0.5
\end{bmatrix}, \\
b^T &= 
   \begin{bmatrix}
   0 & 1
   \end{bmatrix},
\end{split}
&
\begin{split}
    \Delta &= \begin{bmatrix} 0 &   0\\ 0.1 &   -0.1
\end{bmatrix}, \\
 c &= \begin{bmatrix}
   1 & 0
   \end{bmatrix}.
\end{split}
\end{array}
\end{equation}
The optimization problem \eqref{opt2} is solved for $i = 6$ and the resulting Lyapunov parameters $\PP_6$ are used to generate bounds on the impulse response using \eqref{newbound}.  Next, Theorem \ref{thrm4} is employed, and exponential stability guarantees are computed for $\alpha = -0.5,\, 0.15$ and $i = 6$. The exponential and global norm bounds computed in this study are shown in Figure \ref{fig5} in the time domain, along with several sample system impulse responses.
\end{example}

\subsection{Robust Uncertain System Simulation}
As demonstrated in the previous examples, the bound provided in \eqref{newbound}---which is introduced in Theorem \ref{thrm1} and generalised in Theorem \ref{thrm3}---will generally only serve as a good approximation of the system response initially.  Thus, the bound provided in \eqref{newbound} may be too weak to employ in instances where, \emph{e.g.}, long-term system knowledge is needed, and we have attempted to address this concern by providing \emph{e.g.} norm-bounds on the tail of the impulse response (See Examples \ref{example4} and \ref{example5}), and exponential stability bounds (See Theorem \ref{thrm4} and Example \ref{example6}).
As an alternative, we next present a method for approximating $h(t)$ for \eqref{LDI} that uses the {\em difference} between the impulse responses within a family of linear systems. 

We consider 
\begin{equation}\label{diff_impulse}
\begin{split}
\dot{\widetilde{x}} &= 
\begin{bmatrix}
A(t) & 0 \\ 0 & A
\end{bmatrix}
\widetilde{x}+ 
\begin{bmatrix}
b \\ b
\end{bmatrix}u, \vspace{.1in}\\
y & =  
\begin{bmatrix}
c& -c
\end{bmatrix}\widetilde{x},
\end{split}
\end{equation}
and where $A(t)$ satisfies \eqref{acond}. Given a signal $A(t)$, we have that the impulse response of \eqref{diff_impulse} is equal to $h(t) - ce^{At}b$ where $h(t)$ is the impulse response of \eqref{LDI} and is given by \eqref{solution}.  Thus, a new time varying bound on $h(t)$ can be derived straightforwardly from the results presented previously (See Theorem \ref{thrm5}).

Define $A_{+},\, A_{-} \in \R^{2n \times 2n}$, and $\overline{\mathbf{b}}_i,\, \overline{\mathbf{c}}_i^T \in \R^{(2n)^i}$ by
\begin{equation*}
    A_{+} = \begin{bmatrix}
    A + \Delta & 0 \\ 0 & A
    \end{bmatrix},\qquad A_{-} = \begin{bmatrix}
    A-\Delta & 0 \\ 0 & A
    \end{bmatrix},
\end{equation*}
\begin{equation*}
        \overline{\mathbf{b}}_i = (\otimes^{i} \begin{bmatrix} 1 & 1
        \end{bmatrix})^T \otimes \mathbf{b}, \qquad 
        \overline{\mathbf{c}}_i = (\otimes^{i} \begin{bmatrix} 1 & -1
        \end{bmatrix})\otimes \mathbf{c}.
\end{equation*}

\begin{theorem}\label{thrm5}
For $\alpha \in \R$, if $\PP_i \in S_{++}^{(2n)^i}$ satisfies \eqref{metalyap_cond} for $A_{+} + \alpha I_{2n}$ and $A_{-} + \alpha I_{2n}$ at the $i^{\rm th}$ level, then $|h(t) - ce^{At}b|\; \leq e^{-\alpha t}\overline{h}$ for all $t$ where $\overline{h}$ is given by 
\begin{equation}
    \overline{h} = (\overline{\mathbf{c}}_i\PP^{-1}_{i} \overline{\mathbf{c}}_i^T)^{1/(2i)}
    (\overline{\mathbf{b}}_i^T\PP_{i} \overline{\mathbf{b}}_i)^{1/(2i)}.
\end{equation}
Moreover, the parameter $\PP_i$ which minimizes $\overline{h}$ can be computed with \eqref{convexopt}, where $\PP_i$ is understood to satisfy \eqref{metalyap_cond} for both $A_{+} + \alpha I_{2n}$ and $A_{-} + \alpha I_{2n}$ at the $i^{\rm th}$ level.
\end{theorem}

The application of Theorem \ref{thrm5} is demonstrated through a case study in the following section.

\section{Numerical Example}
In this study we consider the uncertain linear system \eqref{LDI} previously introduced  in Example \ref{example6}, and restated here: we consider \eqref{LDI} with $x\in \R^2$ and
\begin{equation}
\begin{array}{ll}
\begin{split}
A &= \begin{bmatrix} 0 &   1\\ -0.6 &   -0.5
\end{bmatrix}, \\
b^T &= 
   \begin{bmatrix}
   0 & 1
   \end{bmatrix},
\end{split}
&
\begin{split}
\Delta &= \begin{bmatrix} 0 &   0\\ 0.1 &   -0.1
\end{bmatrix},\\ 
   c &= \begin{bmatrix}
   1 & 0
   \end{bmatrix}.
 \end{split}
\end{array}
\end{equation}

\begin{figure}[t]
    \input{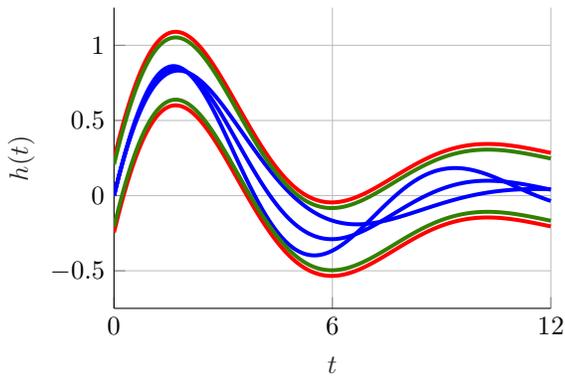}\\
	\caption{Three system simulations are conducted, and the impulse responses of each is plotted in blue. Two envelope bounds on $h(t)$ are computed using Theorem \ref{thrm5} with $\alpha = 0$ and $i=1,\,3$, and these envelope bounds are shown in red and green, respectively.}
    \label{fig6}
\end{figure}

We first consider the case where $\alpha = 0$. We compute $\PP_1$ and $\PP_3$ that satisfies the hypothesis of Theorem \ref{thrm5} using \eqref{convexopt}. By applying Theorem \ref{thrm5}, we compute an envelope centered at $ce^{At}b$ which contains the impulse response of \eqref{LDI}. The bounds computed in this study are shown in Figure \ref{fig6}, plotted in the time domain.

\begin{figure}[t]
    \begin{subfigure}{0.48\textwidth}
%
%
\definecolor{mycolor1}{rgb}{0.20000,0.50000,0.00000}%
\begin{tikzpicture}

\begin{axis}[%
width=6cm,
height=4cm,
at={(0cm,0cm)},
scale only axis,
xmin=0,
xmax=0.2,
xtick={  0, 0.1, 0.2},
xlabel style={font=\color{white!15!black}},
xlabel={$t$},
ymin=-0.1,
ymax=0.3,
ytick={-0.1,    0,  0.1,  0.2,  0.3},
ylabel style={font=\color{white!15!black}},
ylabel={$h(t)$},
axis background/.style={fill=white},
axis x line*=bottom,
axis y line*=left,
xmajorgrids,
ymajorgrids
]
\addplot [color=blue, line width=1.5pt, forget plot]
  table[row sep=crcr]{%
0	0\\
0.01	0.00997494186456004\\
0.02	0.0198995364992426\\
0.03	0.0297734410254073\\
0.04	0.0395963173083285\\
0.05	0.0493678319537726\\
0.06	0.0590876563043078\\
0.07	0.0687554664353501\\
0.08	0.0783709431509441\\
0.09	0.0879337719792835\\
0.1	0.09744364316797\\
0.11	0.106900251679014\\
0.12	0.116303297183578\\
0.13	0.125652484056463\\
0.14	0.134947521370346\\
0.15	0.144188122889755\\
0.16	0.153374007064805\\
0.17	0.162504897024677\\
0.18	0.171580520570852\\
0.19	0.1806006101701\\
0.2	0.189564902947219\\
};
\addplot [color=blue, line width=1.5pt, forget plot]
  table[row sep=crcr]{%
0	0\\
0.01	0.00996997682653287\\
0.02	0.0198798158890475\\
0.03	0.0297293829274226\\
0.04	0.0395185474892517\\
0.05	0.0492471829135682\\
0.06	0.0589151663144791\\
0.07	0.0685223785647081\\
0.08	0.0780687042790502\\
0.09	0.0875540317977379\\
0.1	0.0969782531697225\\
0.11	0.106341264135869\\
0.12	0.115642964112069\\
0.13	0.12488325617227\\
0.14	0.134062047031425\\
0.15	0.143179247028363\\
0.16	0.152234770108577\\
0.17	0.161228533806943\\
0.18	0.170160459230357\\
0.19	0.179030471040301\\
0.2	0.187838497435332\\
};
\addplot [color=blue, line width=1.5pt, forget plot]
  table[row sep=crcr]{%
0	0\\
0.01	0.00997991020681575\\
0.02	0.0199192833114188\\
0.03	0.0298175867759449\\
0.04	0.0396742930575429\\
0.05	0.049488879625288\\
0.06	0.0592608289766786\\
0.07	0.068989628653715\\
0.08	0.078674771258562\\
0.09	0.0883157544687953\\
0.1	0.0979120810522328\\
0.11	0.107463258881351\\
0.12	0.116968800947285\\
0.13	0.126428225373421\\
0.14	0.135841055428569\\
0.15	0.145206819539727\\
0.16	0.154525051304435\\
0.17	0.163795289502715\\
0.18	0.173017078108603\\
0.19	0.182189966301275\\
0.2	0.191313508475761\\
};
\addplot [color=red, line width=1.5pt, forget plot]
  table[row sep=crcr]{%
0	0.0717841341046964\\
0.01	0.0819387608169921\\
0.02	0.0920434900735129\\
0.03	0.10209798012146\\
0.04	0.112101893954769\\
0.05	0.122054899310691\\
0.06	0.131956668666112\\
0.07	0.141806879233604\\
0.08	0.151605212957216\\
0.09	0.161351356507997\\
0.1	0.171045001279269\\
0.11	0.180685843381625\\
0.12	0.190273583637689\\
0.13	0.199807927576607\\
0.14	0.209288585428284\\
0.15	0.218715272117381\\
0.16	0.228087707257044\\
0.17	0.237405615142399\\
0.18	0.246668724743788\\
0.19	0.25587676969977\\
0.2	0.265029488309867\\
};
\addplot [color=red, line width=1.5pt, forget plot]
  table[row sep=crcr]{%
0	-0.0717841341046964\\
0.01	-0.061988877087872\\
0.02	-0.0522444170750277\\
0.03	-0.042551098070646\\
0.04	-0.0329092593381125\\
0.05	-0.0233192354031462\\
0.06	-0.0137813560574963\\
0.07	-0.00429594636290394\\
0.08	0.00513667334467242\\
0.09	0.0145161874505695\\
0.1	0.0238422850566714\\
0.11	0.033114659976403\\
0.12	0.042333010729466\\
0.13	0.0514970405363197\\
0.14	0.0606064573124068\\
0.15	0.0696609736621281\\
0.16	0.0786603068725654\\
0.17	0.0876041789069554\\
0.18	0.0964923163979164\\
0.19	0.105324450640429\\
0.2	0.114100317584571\\
};
\addplot [color=orange, line width=1.5pt, forget plot]
  table[row sep=crcr]{%
0	0.0362058315950043\\
0.01	0.0463622559456662\\
0.02	0.0564692427511979\\
0.03	0.0665264536927727\\
0.04	0.0765335552183349\\
0.05	0.0864902185392915\\
0.06	0.0963961196269375\\
0.07	0.106250939208617\\
0.08	0.11605436276362\\
0.09	0.125806080518821\\
0.1	0.135505787444056\\
0.11	0.145153183247237\\
0.12	0.15474797236922\\
0.13	0.164289863978406\\
0.14	0.1737785719651\\
0.15	0.183213814935608\\
0.16	0.19259531620609\\
0.17	0.201922803796166\\
0.18	0.211196010422268\\
0.19	0.220414673490751\\
0.2	0.229578535090762\\
};
\addplot [color=orange, line width=1.5pt, forget plot]
  table[row sep=crcr]{%
0	-0.0362058315950043\\
0.01	-0.0264123722165462\\
0.02	-0.0166701697527127\\
0.03	-0.00697957164195815\\
0.04	0.00265907939832209\\
0.05	0.0122454453682536\\
0.06	0.0217791929816782\\
0.07	0.0312599936620835\\
0.08	0.0406875235382684\\
0.09	0.0500614634397457\\
0.1	0.0593814988918841\\
0.11	0.0686473201107907\\
0.12	0.0778586219979356\\
0.13	0.0870151041345201\\
0.14	0.0961164707755911\\
0.15	0.105162430843901\\
0.16	0.114152697923519\\
0.17	0.123086990253188\\
0.18	0.131965030719437\\
0.19	0.140786546849449\\
0.2	0.149551270803676\\
};
\addplot [color=mycolor1, line width=1.5pt, forget plot]
  table[row sep=crcr]{%
0	0.0283670645296365\\
0.01	0.0386271001324068\\
0.02	0.0488396537450032\\
0.03	0.0590044112849232\\
0.04	0.0691210637029849\\
0.05	0.079189306982813\\
0.06	0.0892088421400865\\
0.07	0.0991793752215515\\
0.08	0.1091006173038\\
0.09	0.118972284491818\\
0.1	0.1287940979173\\
0.11	0.138565783736739\\
0.12	0.148287073129292\\
0.13	0.157957702294414\\
0.14	0.167577412449279\\
0.15	0.177145949825969\\
0.16	0.186663065668459\\
0.17	0.196128516229373\\
0.18	0.20554206276653\\
0.19	0.21490347153928\\
0.2	0.224212513804625\\
};
\addplot [color=mycolor1, line width=1.5pt, forget plot]
  table[row sep=crcr]{%
0	-0.0283670645296365\\
0.01	-0.0186772164032867\\
0.02	-0.00904058074651803\\
0.03	0.000542470765891326\\
0.04	0.0100715709136721\\
0.05	0.0195463569247322\\
0.06	0.0289664704685292\\
0.07	0.0383315576491486\\
0.08	0.0476412689980878\\
0.09	0.0568952594667489\\
0.1	0.0660931884186405\\
0.11	0.075234719621289\\
0.12	0.0843195212378636\\
0.13	0.0933472658185121\\
0.14	0.102317630291413\\
0.15	0.11123029595354\\
0.16	0.12008494846115\\
0.17	0.128881277819981\\
0.18	0.137618978375175\\
0.19	0.146297748800919\\
0.2	0.154917292089814\\
};
\end{axis}
\end{tikzpicture}%
        \caption{Short time scale: as $\alpha \leq 0$ decrease, the envelope bound \eqref{h_bound} approximates the impulse response of \eqref{LDI} with greater accuracy on short time horizons.
        \vspace{.5cm}
        }
        \label{fig7a}
    \end{subfigure}
    ~
    \begin{subfigure}{0.48\textwidth}
        \input{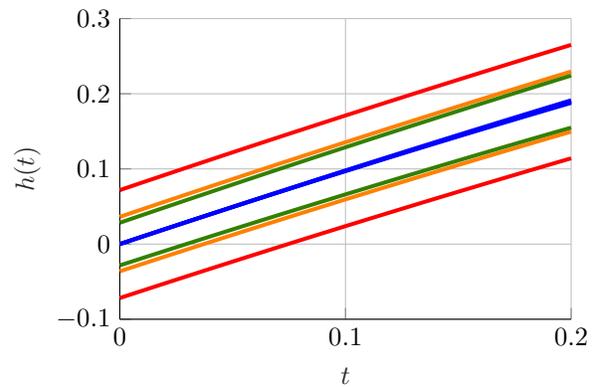}
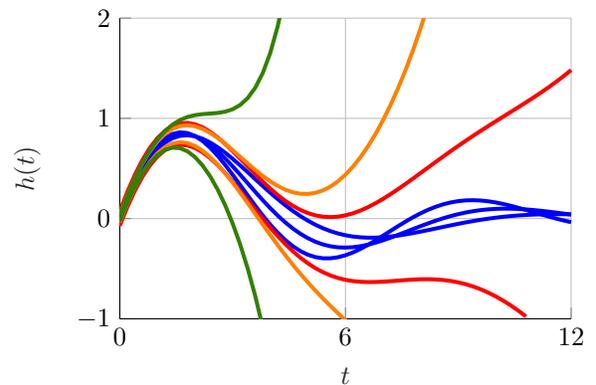
        \caption{Long time scale: as $\alpha \leq 0$ increases, the envelope bound \eqref{h_bound} approximates the impulse response of \eqref{LDI} with greater accuracy on long time horizons.}
		\label{fig7b}
    \end{subfigure}
    \caption{  
    Three system simulations are conducted and the impulse response of each is plotted in blue. Three conference envelopes are formed by applying the procedure detailed in Theorem \ref{thrm5} at level $i = 2$.  We compute $\PP_2 \in \R^{16}$ using \eqref{convexopt} for $\alpha = -0.25,\, -0.5,\, -1$, and the resulting envelope bounds \eqref{h_bound} are shown in red, orange and green, respectively.
    }
    \label{fig7}
\end{figure}

We next consider the case where $\alpha < 0$. In this case, the envelope
\begin{equation}\label{h_bound}
    |h(t) - ce^{At}b|\; \leq e^{-\alpha t}\overline{h},
\end{equation}
that bounds the impulse response $h(t)$ will grow with time. We demonstrate this assertion by computing $\PP_2 \in \R^{(2n)^2}$ that satisfies the hypothesis of Theorem \ref{thrm5} for $\alpha = -0.25,\, -0.5,\, -1$, and plotting the resulting convergence envelopes (See Figure \ref{fig7}).
Note that as $\alpha$ decreases the envelope bound \eqref{h_bound} will approximate the initial system behavior with greater accuracy; however, the long-term accuracy is better achieved with higher values $\alpha$.

\begin{figure}[t]
    \begin{subfigure}{0.48\textwidth}
        \input{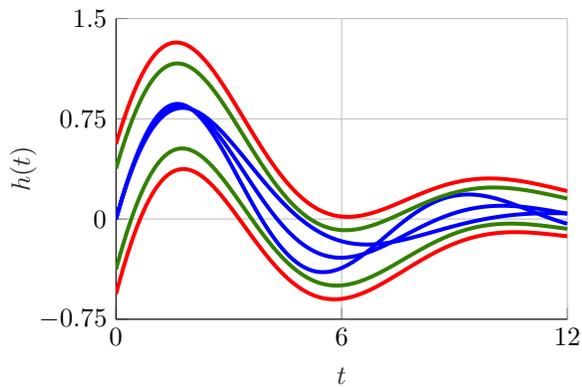}
        \caption{Three envelope bounds are formed by applying the procedure detailed in Theorem \ref{thrm5} at levels $i = 1,\, 3$ with $\alpha = 0.1$. These bounds are shown in red and green, respectively. 
        \vspace{.5cm}
        }
        \label{fig8a}
    \end{subfigure}
    ~
    \begin{subfigure}{0.48\textwidth}
        \input{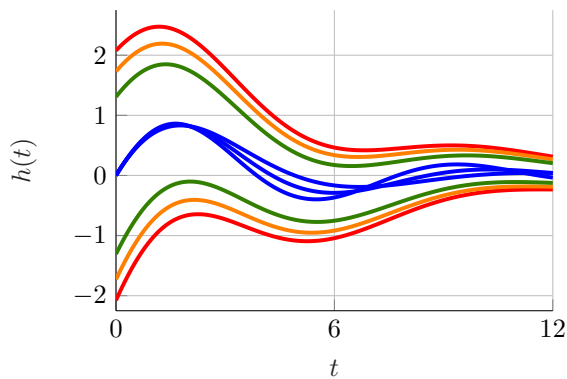}
        \caption{Three envelope bounds are formed by applying the procedure detailed in Theorem \ref{thrm5}. The $i = 1^{\rm st}$ order envelope bound formed by solving \eqref{convexopt} with $\alpha = 0.156$ is shown in red. The $i = 2^{\rm nd}$ order envelope bound formed by solving \eqref{convexopt} with $\alpha = 0.169$ is shown in orange.
        The $i = 3^{\rm rd}$ order envelope bound formed by solving \eqref{convexopt} with $\alpha = 0.173$ is shown in green.}
		\label{fig8b}
    \end{subfigure}
    \caption{  
    Three system simulations are conducted and the impulse response of each is plotted in blue. Convergence envelopes are formed by applying the procedure detailed in Theorem \ref{thrm5} at level $i = 1,\,2,\,3$ for $\alpha \geq 0$.
    }
    \label{fig8}
\end{figure}

Finally, we consider the case where $\alpha > 0$ and, in this case, the resulting envelope \eqref{h_bound} will shrink with time. Moreover, this bound will converge more quickly when $\alpha$ is large and this bound will increase in accuracy as the order of the search $i$ increases. 
To demonstrate this assertion we compute $\PP_i$ that satisfies the hypothesis of Theorem \ref{thrm5} for $\alpha = 0.1$ at the levels $i = 1,\, 3$.  The resulting convergence envelopes are shown in Figure \ref{fig8a}; note that as $i$ increases, the bound \eqref{h_bound} approximates the true maximum impulse response of \eqref{LDI} with greater accuracy.  Additionally, we find that as the order of the search $i$ increases, higher $\alpha$ values are possible.  For instance, when searching for a quadratic Lyapunov parameter $\PP_1$ that satisfies the hypothesis of Theorem \ref{thrm5}, the optimization problem \eqref{convexopt} is solvable only when $\alpha \leq 0.156$.  However, at the $i = 2$ level the optimization problem \eqref{convexopt} is solvable for $\alpha \leq 0.169$, and at the $i = 3$ level the optimization problem \eqref{convexopt} is solvable for $\alpha \leq 0.173$.  The envelope bounds derived from these maximum $\alpha$ parameters are shown in Figure \ref{fig8b}.

\section{Conclusion}
This work demonstrates how the more general class of homogeneous polynomial Lyapunov functions can be used to approximate point-wise-in-time behavior for LTV systems, and we particularly study the impulse and step response of these systems.  Our findings rely on the recent observation that the search for homogeneous polynomial Lyapunov functions for LTV systems can be recast as a search for quadratic Lyapunov functions for a related hierarchy of time-varying Lyapunov differential equations; thus, performance guarantees for LTV systems are attainable without heavy computation.  Numerous examples are provided to demonstrate the findings of this work.
\bibliography{Bibliography}
\bibliographystyle{ieeetr}

\end{document}